# SECURE OUTSOURCED CALCULATIONS WITH HOMOMORPHIC ENCRYPTION

Qi Wang, Dehua Zhou and Yanling Li

Department of Computer Science, Jinan University, China

## ABSTRACT

*With the rapid development of cloud computing, the privacy security incidents occur frequently, especially data security issues. Cloud users would like to upload their sensitive information to cloud service providers in encrypted form rather than the raw data, and to prevent the misuse of data. The main challenge is to securely process or analyze these encrypted data without disclosing any useful information, and to achieve the rights management efficiently. In this paper, we propose the encrypted data processing protocols for cloud computing by utilizing additively homomorphic encryption and proxy cryptography. For the traditional homomorphic encryption schemes with many limitations, which are not suitable for cloud computing applications. We simulate a cloud computing scenario with flexible access control and extend the original homomorphic cryptosystem to suit our scenario by supporting various arithmetical calculations. We also prove the correctness and security of our protocols, and analyze the advantages and performance by comparing with some latest works.*

## 1. INTRODUCTION

Cloud computing is a style of computing in which dynamically scalable and often virtualized resources are provided as a service over Internet, it describes a type of outsourcing of computer services. With the rapid development of cloud computing, there are an increasing number of cloud-based services, such as *Infrastructure-as-a-Service* (*IaaS*), *Platform-as-a-Service* (*PaaS*) and *Software-as-a-Service* (*SaaS*). Under the cloud computing architecture, information is permanently stored in servers on the Internet and cached temporarily on clients [1], users and service providers are separated, data owners and custodians are separated, so the data owners do not have the full control over their own private data, the service provider can access the data that is in the cloud at any time. The cloud providers may share private information with third parties without the permission of the data owners, it is inevitably to incur some new privacy issues. At present, the urgent problem is to study how to protect user privacy efficiently under the cloud computing. To the best of our knowledge, a promising solution to solve such privacy issues is to encrypt the data and then operate on these data in encrypted form. There are lots of schemes have been proposed with various techniques: for instance, encrypted data processing based on secure multiparty computation (SMC) such as the *sharemind* [2], and based on homomorphic encryption (e.g., fully homomorphic encryption (FHE) or partial homomorphic encryption (PHE)) such as [3,4,5,6]. However, the SMC-based scheme usually needs at least two non-colluding cloud servers, for the *sharemind*, it requires at least three servers to complete the secure computations, which is not so convenient for data management. For the FHE-based scheme, it is easily to support various arithmetic computations (e.g., addition and multiplication) over ciphertexts, but is not practical because of the complexity. And the PHE-based scheme which has lower computational complexity but with limited computations (e.g., only support addition or multiplication), there were many previous works [5,6,7] extending PHE-based scheme to support multiple types of computations (e.g., addition, multiplication, exponentiation and natural logarithm), and a boosting linearly-homomorphic encryption scheme [8] has been proposed in 2015, which is capable of evaluating degree-2 computations on ciphertexts by utilizing linearly-homomorphic encryption scheme. But they are not so efficient with too many interactions or with a weaker security setting consideration. It is also a challenge allowing the data requesters to request most of the calculation tasks while not disclosing the original data or mid-result processed by cloud, namely achieving secure data access control in cloud computing [9]. The cloud server should only provide computing service for the delegatee, which is an efficient way to prevent data misuse. In 1998, Blaze, Bleumer and Strauss (BBS) [10] proposed the notion "atomic proxy cryptography", which utilized a semi-honest proxy to convert ciphertexts for Alice into ciphertexts for Bob without seeing the plaintext. Then Dodis and Ivan [11] implemented proxy encryption by dividing the user's secret key into two components, they proposed the unidirectional proxy encryption for ElGamal, RSA and an IBE scheme, but the drawback is that the delegatee needs to store extra secrets for delegate decryption, it is difficult for delegatee to manage the keys. And followed by some pairing based proxy re-encryption schemes have been proposed, such as [12,13], which have been used to many applications. In this paper, we utilize the original BBS's idea, and construct the re-encryption protocol for our scenario.

Most of existing schemes are hard to meet the requirements of various real applications, in this paper, we aim to design a non-interactive privacy-preserving cloud computing system with scalable data access control, which utilizes the partial homomorphic cryptosystem (e.g., the BCP cryptosystem [14]) and the proxy cryptography. The main contributions of our paper can be summarized as below:

(1) We design a complete privacy-preserving cloud computing system with flexible access control.

(2) Our scheme is non-interactive, it reduces the communication overhead efficiently, which is simple and feasible in practical applications.

(3) For the data security, all the raw data and processed data (including intermediate and final results) are not disclosed, only known by the specific person.

The rest of this paper is organized as follows. In section 2, we discuss the related work about encrypted data processing. Section 3 is dedicated to some preliminaries used in our work. Then we give the problem formulation in section 4. And the procedure of our scheme is described in section 5. Section 6 shows the security analysis. And the system evaluation will be given in section 7. Followed by conclusion and future work in section 8.

## 2. RELATED WORK

There are many researches about privacy-preserving data processing with homomorphic encryption, we can divide them into two types: the first type with fully homomorphic encryption (FHE), and the second type with partial homomorphic encryption (PHE). In previous FHE-based works, it is not so efficient for practical because of the storage consumption and computation cost are still too heavy. There are many schemes to support arbitrary computations over encrypted data, for instance, Gentry et al. proposed many schemes with ability to process arbitrary arithmetic computations [3,15], but impractical because of the high computation overhead. In 2012, López-Alt et al. [16] proposed the notion of "*On-the-Fly*" secure multiparty computation on the cloud, it is a multi-key fully homomorphic encryption scheme that allows calculation over data encrypted by different public keys, but still having the efficiency shortcomings as other FHE schemes and relying on an interactive decryption phase, which also leads to heavy communication overhead. And until now, many scholars are still studying how to improve the efficiency of FHE.

In 2013, Anday et al. [17] proposed a privacy-preserving data aggregation scheme based on the revised Paillier's cryptosystem, they divided the decryption key into two parts and distributed them to two different parties. This method reduced the risk of decrypting the raw data directly, but their scheme does not allow multiparty access to the ciphertexts processing result. In the same year, Peter et al. [5] proposed an efficient outsourcing multiparty computation system under multiple keys based on an additively homomorphic encryption (the BCP cryptosystem), in their two-servers model, one of the servers known the master secret key, then two servers performed multiple interactions to compute over encrypted data, but this scheme is not flexible for the data access control. In 2016, Liu et al. [6] achieved a toolkit for efficient and privacy-preserving outsourced calculation that called *EPOM*, which realized by using a double trapdoor decryption cryptosystem (the BCP cryptosystem). In the study, their idea is the same as Anday's, the master private key are distributed to two non-colluding servers, it somewhat reduces the risk of private key leakage, but this scheme can only support the multiplication over a small number of input data. Splitting the single private key to different parties is not flexible for encrypted data computing and management, what is more, this way will produce extra communication overhead. Then in 2017, Ding et al. [7] achieved encrypted data processing with homomorphic re-encryption, their system utilizes two non-colluding servers to manage encrypted data and supports access control on encrypted data processing, but the access control supported by two servers is not suita-ble for many scenarios. In Ding's scheme, the two servers should do Diffie-Hellman key exchange algorithm to generate a joint public key, and then broadcast it to the data providers, we will avoid this drawback in our system. Although Ding's scheme avoid some interactions, the computation overhead for each party is still too heavy, especially for the data requester who wishes to directly decrypt the message to get the correct result. What is more, in Ding's system, it implemented the access control using both two servers, to tackle these problems, we present a scenario achieving the access control with only one server (the access control server), and support ciphertext transfer (e.g., transfer the encrypted data to a specific data requester) by using the re-encrypt technique. Our goal is to simulate the practical scenario, and design lightweight protocols for the encrypted data processing.

## 3. PRELIMINARIES

### 3.1. Additively Homomorphic Encryption

There were many additively homomorphic encryption schemes, such as Paillier's cryptosystem [18], Damgard-Jurik's cryptosystem [19] and some variants of ElGamal scheme (e.g., DGK cryptosystem [20]). In this paper, we use the notation "$[m]$" to denote the encryption of $m$. In an additively homomorphic cryptosystem, it always satisfies the following properties:

(1) *Add*. Given two ciphertexts $[a]$ and $[b]$, then $[a + b] = [a] \odot [b] \bmod N^2$ (addition of two messages, the symbol "$\odot$" represents the multiplication of two ciphertexts).

(2) *cMult*. Given an encrypted message $[a]$ and a constant k in clear, then $[a]^k = [ka] \bmod N^2$ (multiplication of an encrypted message by a known constant).

(3) Note that there is a special property: $[a]^{-1} = [a]^{n-1} = [-a] \bmod N^2$.

### 3.2. The BCP Cryptosystem

The public key cryptosystem with a double trapdoor decryption mechanism proposed by Bresson, Catalano and Pointcheval, namely the BCP cryptosystem [14]. In this cryptosystem, it provides two independent decryption mechanisms, the first decryption mechanism performs the decryption algorithm that can successfully decrypt ciphertexts by specific secret key, while the second decryption mechanism using the master secret key can decrypt any given ciphertext successfully. For simplicity, we only describe the decryption algorithm with the weak secret key (the first decryption mechanism) in this paper, for more details about how to use master key to decrypt the ciphertext successfully (the second decryption mechanism), you can refer to the original paper.

As same as other public key cryptosystems, the BCP cryptosystem also contains four main algorithms: *Setup*, *KeyGen* (key generation), *Enc* (encryption)and *Dec* (decryption). This cryptosystem is semantically secure, we briefly recall the construction of it as follows.

*Setup*: This algorithm generates the public parameters by inputting a security parameter $\kappa$. Let $N = pq$ be an RSA modulus with the length of $\kappa$, $g$ is an element of maximal order in $\mathbb{G}$, where $\mathbb{G}$ is the cyclic group of quadratic residues modulo $N^2$. The plaintext space is $\mathbb{Z}_N$, the public parameters $PP = (N, g)$ and the master key $MK = (p, q)$.

*KeyGen*: Given the public parameters $PP$, then randomly choose a secret values $x \in [1, \text{ord}(\mathbb{G})]$, we can compute $h = g^x \bmod N^2$. The public key $pk = (N, g, h)$, while the corresponding private key is $sk = x$.

*Enc*: Given the user's public key and a message $m \in \mathbb{Z}_N$, and choose a random number $r \in \mathbb{Z}_{N^2}$. This algorithm outputs the ciphertext $[m]$ as

$$[m] = (A, B) = (g^r \bmod N^2, \ (1 + mN)h^r \bmod N^2). \tag{1}$$

*Dec*: Given the ciphertext $[m]$ and the user's weak secret key $x$, this algorithm outputs the plaintext $m$ as

$$m = L(B/A^x \bmod N^2), \text{ where } L(u) = \frac{u-1}{N} \ . \tag{2}$$

Next, we assume that the weak private key is $sk = sk_1 + sk_2$, and the corresponding public key is $pk' = (N, g, h')$, where $h' = g^{sk} \bmod N^2$. The encryption algorithm is the same as depicted above but encrypted with $pk'$. We present a *two-phase decryption mechanism* as follows:

*Partial decryption with $sk_1$ (PDec1)*: This algorithm can transfer the original ciphertext into another ciphertext that can be decrypted by $sk_2$, it operates as follows:

$$[m]' = (A', B') = (g^r \bmod N^2, \ (1 + mN)g^{sk_2 r} \bmod N^2), \text{ where } B' = \frac{B}{A^{sk_1}} \bmod N^2 \ . \tag{3}$$

*Partial decryption with $sk_2$ (PDec2)*: This algorithm can directly decrypt the ciphertext from *PDec1* by using $sk_2$ as follows:

$$m = L(B'/A^{sk_2} \bmod N^2), \text{ where } L(u) = \frac{u-1}{N} \ . \tag{4}$$

# 4. PROBLEM FORMULATION

## 4.1. System Model

In this work, we focus on encrypted data processing under the cloud computing. As you know, in the normal cloud computing environment, the service provider can access the data that is in the cloud at any time, this poses serious privacy concerns. To protect data in terms of confidentiality and privacy from unauthorized users, we propose a practical scheme for encrypted data processing under cloud computing. Our system comprises four types of entities, as shown in Figure 1.

(1) *Cloud Service Provider (CSP)*. This entity mainly stores the encrypted data from data providers and provides some homomorphic computation service. We can think of it as a storage cloud with computing capabilities.

(2) *Access Control Server (ACS).* The entity provides service of secure data computation with data access control for the target user, and performs the ciphertext transformation. In our scenario, there could exist several ACSs.

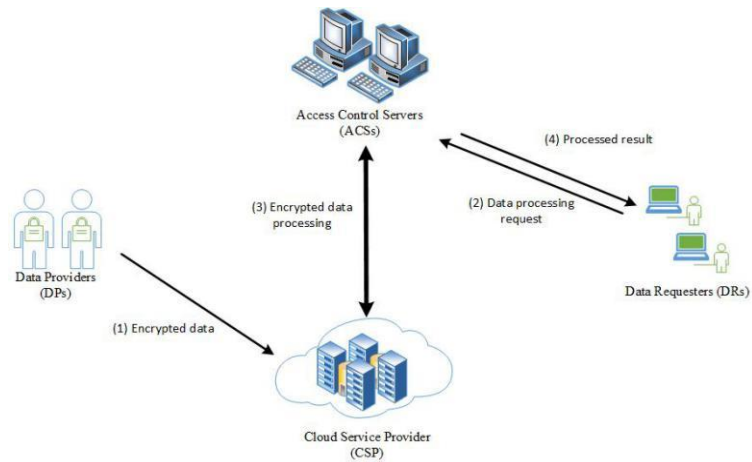

Figure 1. System overview

(3) *Data Providers (DPs).* There are several data providers encrypting their private data and uploading these ciphertexts to the CSP. These encrypted data stored in the cloud will be used for further computation and analysis.

(4) *Data Requesters (DRs).* These parties are the data consumers that acquire the result of specific data processing. We point out that a DR could also be a DP. Each DR could choose one of the ACSs he trusts, and he only communicates with ACS (e.g., puts a data processing request and then receives the processed result).

In the beginning, each DP will choose an ACS he trusts, then encrypts his private data using the combined public key of ACS and CSP, and then uploads these ciphertexts to CSP. When DR puts a request to the ACS he trusts, the ACS will generate a re-encryption key for him, then sends this key and the request to CSP, so the CSP will cooperate with ACSs to process the outsourced encrypted data which satisfy the demands of the DR. At the end, the chosen ACS will return the final encrypted results to DR who can decrypt to get the plain processing results. In our system, we design the non-interactive protocols to process encrypted data.

## 4.2. Privacy Requirements

Privacy preserving is crucial for the success of our outsourced data processing service. In our system, we consider both the CSP and the ACS are semi-honest (e.g., they will execute the protocols honestly but curious about sensitive data), and they do not collude with each other. In order to prevent each party from information leakage, our system should satisfy the following privacy requirements.

(1) DP's privacy: The DP's private data stored in the cloud should be permanently confidential, it cannot be directly exposed to any other parties, and we must ensure these private data will not be used directly by the CSP.

(2) DR's privacy: The DR puts the encrypted data processing request (encrypted by CSP's public key) to the ACS he trusts, but the ACS should not disclose DR's identity to other parties. And the processed results should only be decrypted by the DR.

## 4.3. Design Goals

In order to achieve an efficient secure system which fulfills the aforementioned scenario and privacy requirements, our scheme should consider the security and performance overhead (mainly the communication cost and computation cost) as follows.

(1) *Correctness and security*. To achieve the privacy-preserving cloud computing system, it must guarantee the correctness of our protocols and the privacy requirements as mentioned before.

(2) *Lightweight*. We aim to design a lightweight system with minimum communication and computation overhead. For communication cost, we must minimize the communication cost for each party, especially for DR, he just needs to put requests and then receives the processed results. And for computation cost, in order to save the computing resource, we should better design the arithmetic operations over encrypted data with light-weight calculation.

# 5 .THE PROCEDURES OF OUR SCHEME

## 5.1. General Description

We propose the privacy-preserving data processing scheme under the cloud computing environment, it utilizes the BCP cryptosystem [14] to realize our system, the procedures are depicted as follows:

*Init*: This procedure initializes the system parameters with the *Setup* algorithm of BCP cryptosystem, and broadcasts the public parameters $PP = (N, g)$ to all parties.

*KeyGen*: After all participants received the parameters, they choose random private number as their secret key, respectively. For instance, the CSP chooses $a$, the ACS chooses b, the DR chooses c. So the public key of each corresponding party is $g^a$, $g^b$ and $g^c \ mod \ N^2$. It is easily to compute the joint public key of ACS and CSP as $PK = g^{a+b} \ mod \ N^2$.

*Re-KeyGen*: This algorithm is executed by ACS. When DR puts the request to ACS, the ACS will generate a re-encryption key for DR, which computed as $rk_{ACS \rightarrow DR} = g^{c/b} \ mod \ N^2$. Then the $rk$ and the request will be sent to CSP.

*Enc*: This algorithm is executed by DPs. Using the combined public key PK, DPs encrypt their private data and then upload the ciphertexts to the CSP. For example, we assume a message $m \in \mathbb{Z}_N$, then choose a random number $r \in \mathbb{Z}_{N^2}$, so the ciphertext $[m]$ is computed as:

$$[m] = (A, B) = (g^r \bmod N^2, (1 + mN)h^r \bmod N^2) . \qquad (5)$$

*Partial-Dec* (*PDec*): This algorithm is executed by CSP. Using CSP's secret key $a$, it can get a fresh ciphertext which can be decrypted by ACS. The fresh ciphertext is:

$$[m]' = (A', B') = \left(g^r \bmod N^2, \ (1 + mN)g^{br} \bmod N^2\right), \text{ where } B' = \frac{B}{A^a} \bmod N^2 . \quad (6)$$

Then CSP adds noise $\delta$ to $[m]'$ and get $[m + \delta]_{ACS}$, in the meanwhile he encrypts $\delta$ with $rk_{ACS \to DR}$. The CSP sends $[m + \delta]_{ACS}$ and $[\delta]_{rk}$ to ACS.

*Re-Enc*: This step performed by ACS. ACS uses his secret key b to decrypt $[m + \delta]_{ACS}$ and transfer $[\delta]_{rk}$ to $[\delta]_{DR}$. After performing this step correctly, the ciphertexts will be switched to $[m]_{DR}$ (a message encrypted by DR's public key).

*Dec*: When DR receives the data from ACS, he can decrypt it with his secret key c to get the processed result.

## 5.2. Encrypted Data processing

We assume that the encrypted data has been stored in the CSP, and the latter operations on ciphertexts do not need the DPs to participate in. So we introduce how to implement two basic operations (addition and multiplication) herein.

***Addition.*** This operation aims to obtain the sum of some raw data: $m = \sum_{i=1}^{n} m_i$.

Step 1 (@ CSP): Due to the additively homomorphic property, the CSP can directly multiply the ciphertexts one by one as following:

$$[m]_{PK} = \prod_{i=1}^{n} [m_i]_{PK} . \qquad (7)$$

Then as described before, the CSP partially decrypts $[m]_{PK}$ and adds noise to it, sends the data package ($[m + \delta]_{ACS}, [\delta]_{rk}$) to ACS.

Step 2 (@ ACS): ACS utilizes his secret key b to process the encrypted data, switch the ciphertexts to $[m]_{DR}$, which is the encrypted data processed result can be decrypted by DR who puts the request.

Step 3 (@ DR): Upon receiving the encrypted data from ACS, the DR decrypts it to get the clear processed result.

***Multiplication.*** This operation aims to obtain the product of some non-zero raw data: $m = \prod_{i=1}^{n} m_i$. For ease of presentation, we give the instance of multiplying two messages here.

Step 1 (@ CSP): The CSP first chooses two random numbers $r_1$ and $r_2$, and computes $r_3 = (r_1 * r_2)^{-1}$. Then he performs the *PDec1* algorithm to get $[m_1]_{ACS}$ and $[m_2]_{ACS}$, computes $[m_1]_{ACS}{}^{r_1}$ and $[m_2]_{ACS}{}^{r_2}$, sends the data package ($[m_1 r_1]_{ACS}, [m_2 r_2]_{ACS}, [r_3]_{rk}$) to ACS.

Step 2 (@ ACS): Upon receiving the data package, the ACS decrypts $[m_1 r_1]_{ACS}$ and $[m_2 r_2]_{ACS}$ to get the message with "noise", and computes $m_1 r_1 * m_2 r_2$. Then, the ACS switches $[r_3]_{rk}$ to $[r_3]_{DR}$ using his secret key b. Finally, the ACS computes $[r_3]_{DR}{}^{m_1 r_1 * m_2 r_2}$ to get the final result $[m_1 m_2]_{DR}$.

Step 3 (@ DR): The DR can obtain the correct result after decrypting the data from ACS.

# 6. SECURITY ANALYSIS

As described in the previous section, the correctness of our protocols is obvious, so we focus on the security of our system in this section. Our system aims to achieve the privacy of user data in cloud, the intermediate and final results security under the semi-honest model. And we assume that there is no collusion between the CSP and ACS. We first introduce the semantic security of BCP cryptosystem.

**Theorem 1.** *If Decisional Diffie-Hellman Assumption in $\mathbb{Z}_{N^2}^*$ holds, then the BCP cryptosystem is semantically secure.*

*Proof.* We say the Diffie-Hellman computational problem is hard, if for every probabilistic polynomial time algorithm $\mathcal{A}$, there exists a negligible function $negl()$ such that for sufficiently large $\ell$.

$$\Pr\left[\mathcal{A}(N, X, Y, Z_b) = b \;\middle|\; \begin{array}{l} p, q \leftarrow SP\left(\frac{\ell}{2}\right); N = pq; g \leftarrow \mathbb{G}; \\ x, y, z \leftarrow [1, ord(\mathbb{G})]; X = g^x \bmod N^2; \\ Y = g^y \bmod N^2; Z_0 = g^z \bmod N^2; \\ Z_1 = g^{xy} \bmod N^2; b \leftarrow \{0,1\}; \end{array}\right] - \frac{1}{2} = negl(\ell) \quad .(8)$$

Given a quadruple $(g, g^a, g^b, g^c)$, to use $\mathcal{A}$ to decide whether $c = ab \bmod ord(\mathbb{G})$ or not. The public key is first set as $(N, g, h)$, where $h = g^a$. Once the adversary has chosen the messages $m_0$ and $m_1$, we flip a coin $d$ and then encrypt the message $m_d$ as follows:

$$E(m_d) = (A, B) = \left(g^b \bmod N^2, (1 + m_d N)g^c \bmod N^2\right) \quad . \tag{9}$$

If this quadruple is a Diffie-Hellman quadruple (i.e., $c = ab$), the above is a valid encryption of $m_d$ and $\mathcal{A}$ will give the correct response with non-negligible advantage. If it is not a Diffie-Hellamn quadruple, we claim that even a polynomial unbounded adversary gains no extra information from $E(m_d)$ in a strong information-theoretic sense. You can refer the original BCP paper [14] for the detail.

**Corollary 1** *(Two-phase decryption security). The two-phase decryption mechanism of the BCP cryptosystem is semantically secure based on the hardness of DDH assumption over $\mathbb{Z}_{N^2}^*$.*

*Proof.* The Shamir secret sharing scheme [21] proposed in 1979, which is information theoretic secure can guarantee the privacy of divided private key. The private message is randomly split into two shares in a way that any less than two shares cannot recover the original message (i.e., *the (2, 2) Shamir threshold secret sharing scheme*). And the adversary can only recover the original plaintext by both two shares of partial decrypted ciphertexts.

**Lemma 1***(Privacy of data). While the two-phase decryption BCP cryptosystem is semantically secure, without collusion, our system always protects the security of the data (e.g., user data and processed results).*

*User data privacy*. We claim the privacy of user data stored in the cloud server here. The original private data are encrypted by data owners with the joint public keys of CSP and ACS, and stored in the CSP who can only decrypt the encrypted data partially. As we assumed that the CSP and the ACS do not collude with each other, the user data in the cloud will be permanently confidential.

*Processed result privacy*. We show the privacy preserving of the processed result. The CSP do calculations over ciphertexts, and he has no ability to decrypt it completely. So he decrypts the encrypted result partially with his own secret key, and blinds the result with a random message (namely, the "*noise*"), he sends the blinded result to ACS, although ACS has another half secret key, he only decrypts the message to get the blinded result. While the noise were encrypted with $rk$, no one can decrypt it directly, the ACS processes the encrypted with his own secret key, so it can be transformed to the data encrypted by DR′s public key. Then the ACS performs encrypted data calculations to get the final result under the DR′s public key, which will be sent to DR. In our procedures, all the mid/final results are not disclosed, and only the DR who puts the request can get the final result.

What is more, we also protect the identity of data requester, only the access control server (ACS) recognizes DR′s identity. But in Ding′s scheme [7], the access control was implemented by both CSP and ACS.

## 7. EVALUATION AND IMPLEMENTATION

### 7.1. Scheme Evaluation

In this section, we compare our system with some latest work [5,6,7], which were also the similar researches with partial homomorphic encryption (PHE). And there are also some other works based on different techniques, for instance, SMC-based [2,22] and FHE-based [4,23], but we think FHE and SMC will introduce significant communication/computation overhead. We aim to achieve the lightweight scheme, so there should be no extra computations for client, and less interactions in the procedures will be better. For the security concern, using the master key will increase the risk of system, if the master key is disclosed, all these data will not secure. Our scheme overcomes all these drawbacks, and more efficient for the practical.

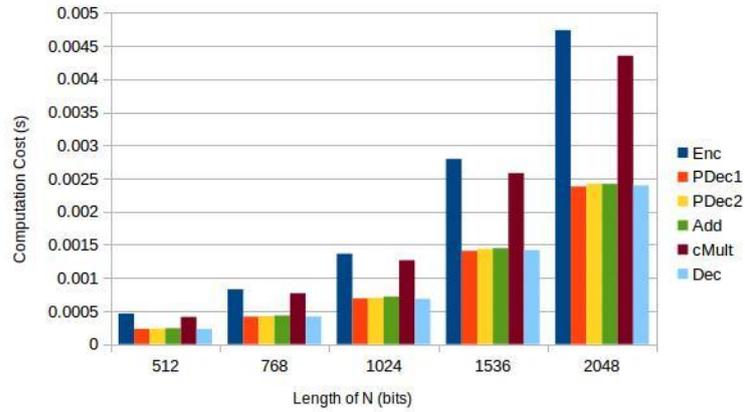

Figure 2. Computation time of each algorithm in BCP with various N

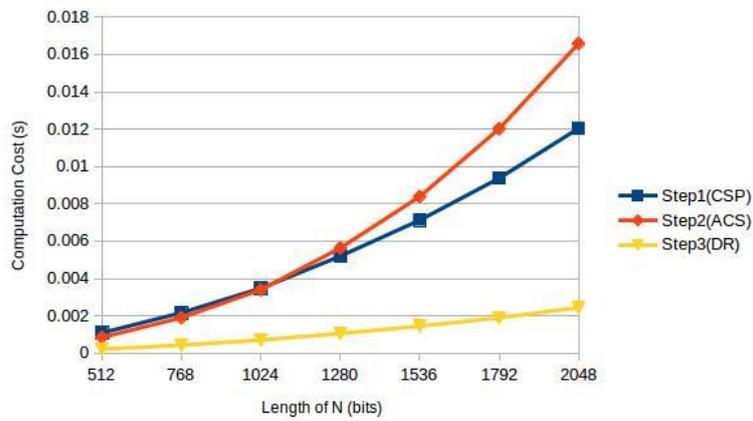

Figure 3. Computation time of addition protocol with various N

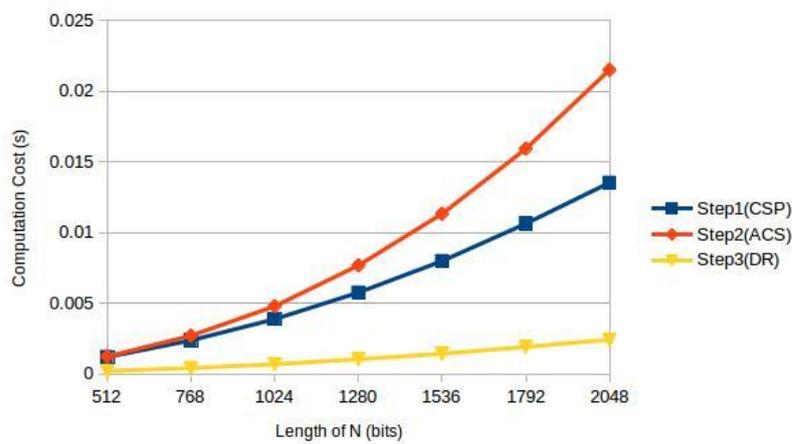

Figure 4. Computation time of multiplication protocol with various N

## 7.2. Experiment Analysis

We have implemented our proposed scheme with Python, the experiments are conducted on Ubuntu 16.04.3 LTS with Intel(R) Core(TM) i7-6700 CPU 3.40GHz and 4GB RAM. To achieve better accuracy, we performed each test 500 times and reported the average value of all results. In our implementation, we tested the influence of different parameters (i.e., the length of secret key, the length of data and the length of N), and we found that the length of the parameter N is significant. So we set other parameters by default, for instance, the length of secret key is 500 bits, the length of data is 200 bits and the length of random number is 500 bits. Then we tested the performance of each algorithm with different length of N (i.e., 512 bits, 768 bits, 1024 bits, 1280 bits and so on).

The performance of each algorithm in BCP cryptosystem was shown in Figure 2, the computation costs affected much by the length of N.And we tested the performance of our two proposed protocols (addition and multiplication) in section 5.2, we analyzed the computation costs of each step, corresponding to three parties: CSP (the step1), ACS (the step2) and DR (the step3). As you can see in Figure 3 and Figure 4, both in addition and multiplication protocol, most of the computation overhead has been transferred to DSP and ACS, the computation costs of DR is little, when he received the result, he can decrypt it in a few milliseconds.

In general, the above tests show that we achieve the lightweight of DR in our scheme, nearly all of the computation overhead has been transferred to CSP and ACS, which is practical in the real cloud computing environments.

## 8. CONCLUSION AND FUTURE WORK

In this paper, we proposed an efficient encrypted data processing scheme with access control. We overcame some drawbacks of existing works, and designed some lightweight protocols which are non-interactive and more flexible for access control. In the future, we aim to construct a more secure system with multi-key homomorphic cryptosystem, which can provide better privacy-preserving service for data providers. What is more, we also want to extend our scheme to a malicious one, that is, even the CSP and ACS collude with each other, there are no sensitive information disclosed.